# Strain-driven stabilization of a room-temperature chiral multiferroic with coupled ferroaxial and ferroelectric order

Guodong Ren[1], Gwan Yeong Jung[2], Huandong Chen[3], Chong Wang[4], Boyang Zhao[3], Rama K. Vasudevan[5], Jordan A. Hachtel[5], Andrew R. Lupini[5], Miaofang Chi[5], Di Xiao[4], Jayakanth Ravichandran[3,6], Rohan Mishra[1,2*]

[1] *Institute of Materials Science and Engineering, Washington University in St. Louis, St. Louis, MO 63130, USA*

[2] *Department of Mechanical Engineering and Material Science, Washington University in St. Louis, St. Louis, MO 63130, USA*

[3] *Mork Family Department of Chemical Engineering and Materials Science, University of Southern California, Los Angeles, CA 90089, USA*

[4] *Department of Materials Science and Engineering, University of Washington, Seattle, WA 98195, USA*

[5] *Center for Nanophase Materials Sciences, Oak Ridge National Laboratory, Oak Ridge, TN 37830, USA*

[6] *Ming Hsieh Department of Electrical Engineering, University of Southern California, Los Angeles, CA 90089, USA*

*Corresponding author: rmishra@wustl.edu

**Abstract**

Noncollinear ferroic materials are sought after as testbeds to explore the intimate connections between topology and symmetry, which result in electronic, optical and magnetic functionalities not observed in collinear ferroic materials. For example, ferroaxial materials have ordered rotational structural distortions that break mirror symmetry and induce chirality. When ferroaxial order is coupled with ferroelectricity arising from a broken inversion symmetry, it offers the prospect of electric-field-control of the ferroaxial distortions and opens up new tunable functionalities. However, chiral multiferroics, especially ones stable at room temperature, are rare. We report the discovery of a strain-stabilized, room-temperature chiral multiferroic phase in single crystals of $BaTiS_3$, a quasi-one-dimensional (1D) hexagonal chalcogenide. Using first-principles calculations, we predict the stabilization of this multiferroic phase having $P6_3$ space group for biaxial tensile strains exceeding 1.5% applied on the basal *ab*-plane of the room temperature $P6_3cm$ phase of $BaTiS_3$. The chiral multiferroic phase is characterized by rotational distortions of select $TiS_6$ octahedra around the long *c*-axis and polar displacement of Ti atoms along the *c*-axis. We used an innovative approach using focused ion beam milling to make appropriately strained samples of BaTiS$_3$. The ferroaxial and ferroelectric distortions, and their domains in *P*6$_3$-BaTiS$_3$ were directly resolved using atomic resolution scanning transmission electron microscopy. Landau-based phenomenological modeling predicts a strong coupling between the ferroelectric and the ferroaxial order making *P*6$_3$-BaTiS$_3$ an attractive test bed for achieving electric-field control of chirality-related phenomena such as circular photo-galvanic current and the Rashba effect.

**Notice:** This manuscript was authored by UT-Battelle, LLC, under Contract No. DE-AC0500OR22725 with the U.S. Department of Energy. The United States Government retains and the publisher, by accepting the article for publication, acknowledges that the United States Government retains a nonexclusive, paid-up, irrevocable, worldwide license to publish or reproduce the published form of this manuscript or allow others to do so, for the United States Government purposes. The Department of Energy will provide public access to these results of federally sponsored research in accordance with the DOE Public Access Plan (http: //energy.gov/downloads/doe-public-access-plan).

**Main Text**

Ferroic materials are characterized by spontaneous symmetry breaking with the order parameter being responsive to external fields. They form important components of modern-day electronics[1-2]. Conventional ferroic materials include ferroelectrics – that undergo inversion symmetry breaking leading to electric polarization, ferromagnets – that undergo time-reversal symmetry breaking leading to spin-polarization or magnetization, and ferroelastics – that undergo symmetry lowering leading to spontaneous strain. Recently, there is a growing interest in ferroic materials having non-collinear order due to their fascinating connections between topology and symmetry, which can enable new electronic, magnetic and photonic functionalities[3-6]. For example, the stabilization of magnetic skyrmions have opened up pathways for low-power information storage[6]. Non-collinear ferroics include ferrotoroidal and ferroaxial materials[3, 7-9]. Ferrotoroids have a toroidal arrangement of magnetic moments and break both space-inversion and time-reversal symmetries[9-10]. Ferroaxial (or ferrorotational) materials have rotational structural distortions that preserve both space-inversion and time-reversal symmetries, but break mirror symmetry[7-9, 11-12]. Ferroaxial materials can potentially demonstrate bulk or planar chirality and lead to such physical phenomena as circular dichroism, optical rotation[8, 13-15], and topological structures such as dipolar skyrmions[16-18]. Dynamic control over these phenomena can enable energy efficient devices for information storage and processing. However, purely ferroaxial materials cannot couple with external fields[19], and hence, external control of chirality is challenging, although few solutions do exist[18, 20-21].

Ferroaxial order, if coupled with either ferroelectric or ferroelastic order, can overcome the above limitation and enable control over chirality using an external field. It can also result in rich physical phenomena such as ferroelectric Rashba effect[22], and tunable circular photogalvanic response effect[23] and Kramer-Weyl fermions[4]. A group theoretical analysis shows that there are 124 unique phase transitions from a high-symmetry parent phase that result in a lower symmetry ferroaxial phase; 64 of those also result in ferroelectricity[19]. Despite the large probability of finding chiral multiferroics with simultaneous ferroelectric and ferroaxial order, there are only a handful of reports of such materials. Examples include hexagonal $BaTiO_3$ below 74 K[22], $Cu_3Nb_2O_8$ below 24 K[12], $RbFe(MoO_4)_2$ below 4 K[23], $CaMn_7O_{12}$ below 90 K[11]; and certain organic crystals including Rochelle salt below 297 K and triglycine sulfate below 321 K[24]. If one ignores the organic crystals due to their incompatibility with the processing conditions of modern electronics, all the reported inorganic materials are stabilized in a chiral ferroic phase much below room temperature.

In this Article, we report the strain-induced stabilization of a chiral multiferroic phase having a strong coupling between the ferroelectric and ferroaxial order parameters in single crystals of a quasi-one-dimensional (1D) chalcogenide $BaTiS_3$. At room temperature, $BaTiS_3$ exists in the achiral $P6_3cm$ phase, which has antipolar displacements of Ti atoms along the *c*-axis[25-26]. Using group theory and first-principles calculations, we predict that a biaxial strain can stabilize $BaTiS_3$ in the $P6_3$ space group — that involves a rotation of $TiS_6$ octahedra about the *c*-axis and polar displacements of Ti atoms along the *c*-axis. Using phenomenological Landau theory-based modeling, we predict a strong coupling between the two order parameters. We devised an innovative focused ion beam (FIB)-based strategy to biaxially strain $BaTiS_3$ single crystals by over 3% — that is outside the range of strains achievable using epitaxy — and stabilize the chiral multiferroic phase at room temperature. We used scanning transmission electron microscopy (STEM)-based imaging and diffraction methods to directly map the ferroelectric and ferroaxial order parameters with atomic resolution, thus confirming the theoretical predictions. We also show the presence of nanoscale chiral domains with sharp domain boundaries. We expect that the room-temperature chiral multiferroic phase in BaTiS$_3$ can enable strong electric-field switching of chirality.

**Theoretical prediction of strain-induced stabilization of a chiral multiferroic phase in BaTiS$_3$**

BaTiS$_3$ is a quasi-1D semiconductor having a small band gap (~0.3 eV)[25-27]. It shows giant optical anisotropy[26-27], and glass-like thermal properties[28]. It also shows multiple electronic and structural phase transitions upon cooling[25]. At room temperature, the global refinement of synchrotron X-ray diffraction of BaTiS$_3$ single crystals leads to an average structural symmetry corresponding to the $P6_3cm$ space group[25, 29]. This structure consists of face-shared TiS$_6$ octahedral chains along the $c$-axis. The TiS$_6$ chains are arranged in a hexagonal manner and are stacked between chains of Ba atoms, as illustrated in Fig. 1(a). $P6_3cm$ is a noncentrosymmetric space group lacking space-inversion symmetry along the $c$-axis, which is caused by the off-centering or polar displacements of Ti atom from the centroid of the TiS$_6$ octahedron. Within a chain, the Ti polar displacements along the $c$-axis are aligned parallel to each other. Between neighboring chains, these displacements are aligned in an antiparallel manner, resulting in an anti-polar ordering. There is a net polarization along the $c$-axis due to unequal upward ($P+$) and downward ($P-$) Ti polar displacements in the unit cell, as illustrated in Fig. 1(a). The Ti atoms are also off-centered along the $ab$-plane. At room temperature, these $ab$-plane displacements are globally disordered, but are ordered locally over a few unit cells[26]. Below 240 K, the $ab$-plane Ti displacements order to form a three-dimensional toroidal dipolar structure with antiparallel dipoles along the $c$-axis and an array of vortex-vortex-antivortex dipolar pattern along the $ab$-plane[29]. Given its susceptibility to form non-collinear dipolar textures at cryogenic temperatures, we selected BaTiS$_3$ to explore whether non-collinear distortions can also be stabilized at room temperature with external perturbations such as strain.

We used group-theoretical tools to search for symmetry-related sub-groups of the room-temperature $P6_3cm$ phase that belong to one of the 65 Sohncke groups and are expected to show chirality[30]. We find that the chiral $P6_3$ sub-group can be accessed with minimal distortions, and it also lacks inversion symmetry. The mirror-symmetry breaking from $P6_3cm$-BaTiS$_3$ can be introduced through a $\Gamma_2$ distortion mode, which results in either clockwise or anticlockwise rotations of the TiS$_6$ octahedral chains lying at the corner of a unit cell, with the rotations being in-phase along a chain. These rotational distortions introduce a ferroaxial order, $A+$ or $A-$ parallel to the 6-fold rotation axis ($c$-axis), as shown schematically in Fig. 1(b-c). Meanwhile, a $\Gamma_1$ distortion mode drives the Ti polar displacements along the $c$-axis to align in a parallel manner between neighboring chains in the $P6_3$ phase, as opposed to their antipolar ordering in the $P6_3cm$ phase. Additionally, the $\Gamma_1$ distortion involves outward displacements of neighboring Ba atoms, causing an expansion of some of the Ba hexagons and leading to the dimerization of Ba atoms, as illustrated in Fig. S1(d-e). Further details of the distortion mode analysis are provided in Supplementary Section 1. Together, these two distortion modes lead to simultaneous ferroelectric and ferroaxial transitions from achiral $P6_3cm$-BaTiS$_3$ to chiral $P6_3$-BaTiS$_3$. The different combinations of chirality ($A+$ or $A-$) and polarization ($P+$ or $P-$) in $P6_3$-BaTiS$_3$ can be expected to result in a multi-domain structure with four degenerate chiral multiferroic states ($A-P-$, $A-P+$, $A+P-$, and $A+P+$), which are related to each other by glide and mirror symmetries.

Next, we used first-principles density-functional theory (DFT) calculations to find the conditions that can activate the $\Gamma_1$ and $\Gamma_2$ distortion modes in achiral $P6_3cm$-BaTiS$_3$ and stabilize the chiral multiferroic $P6_3$-BaTiS$_3$. A geometry optimization was performed with the $\Gamma_1$ and $\Gamma_2$ distortion modes imposed in achiral $P6_3cm$-BaTiS$_3$. The optimized crystal structure was found to be in a local minimum and retained both the ferroelectric and ferroaxial order, as shown in Fig. 1(b-c). Moreover, the optimized $P6_3$-BaTiS$_3$ lattice was found to have larger lattice constants along the *ab* plane with $a = b = 12.03$ Å as compared to $a = b = 11.82$ Å in $P6_3cm$-BaTiS$_3$, which suggests that the chiral multiferroic phase in BaTiS$_3$ can be potentially stabilized using biaxial tensile strain along the *ab*-plane. The calculation details can be found in *Methods* and *Supplementary Section 2*. We use the optimized lattice parameters of $P6_3cm$-BaTiS$_3$ as a reference (i.e., zero strain). At zero strain, $P6_3$-BaTiS$_3$ is energetically unfavorable. However, its energy is lowered with respect to $P6_3cm$-BaTiS$_3$ with increasing biaxial tensile strain along the *ab*-plane. For biaxial strains > 1.5% (tantamount to uniaxial-compressive strain along the *c*-axis of < –0.43%), we find that the $P6_3$-BaTiS$_3$ phase becomes lower in energy than $P6_3cm$-BaTiS$_3$ [Fig. S2(a)], thus providing a pathway to realize the chiral multiferroic phase experimentally. In addition to the rotational distortions of TiS$_6$ octahedra, the optimized $P6_3$-BaTiS$_3$ lattice also displays concomitant *ab*-plane displacements of Ba atoms that lead to three distinct local Ba$_6$TiS$_6$ structural motifs [*Supplementary Section 1* Fig. S1(d-f)].

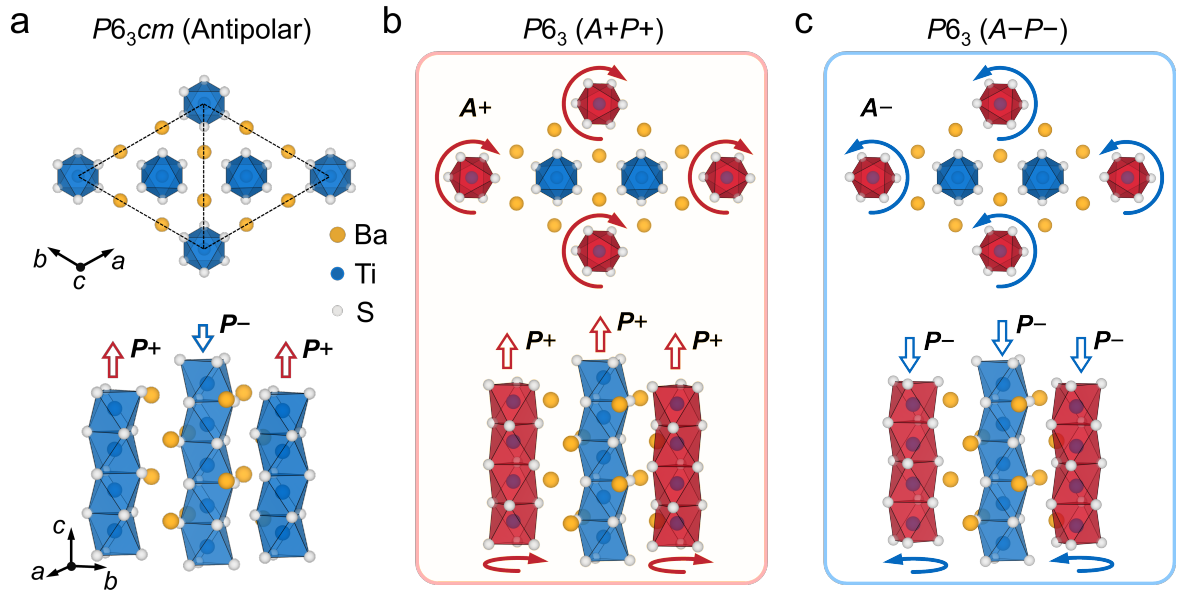

**Figure 1. Atomic models showing chiral multiferroic order in BaTiS$_3$. a.** Achiral $P6_3cm$-BaTiS$_3$ lacks space-inversion symmetry along the *c*-axis due to off-center displacements of Ti atoms within the TiS$_6$ octahedra, as shown in the bottom panel. These Ti displacements have an overall ferrielectric order due to a mixture of parallel and antiparallel alignment of the dipoles, and lead to two net polar states (*P*+ and *P*–) depending on the direction of the displacements. For simplicity, only the positive polar state (*P*+) is shown. The dash lines in the top panel denote the mirror planes including a 6-fold rotation axis. **b, c.** Simultaneous ferroaxial order and ferroelectric order in $P6_3$-BaTiS$_3$. Two (out of the four) domain states, *A*+*P*+ and *A*–*P*–, are shown. In the top-panels, a clockwise rotation (*A*+) of TiS$_6$ octahedra is shown in **b** and an anticlockwise rotation (*A*–) in **c**. Blue and red arrows represent the direction of the rotational distortions. The rotated TiS$_6$ octahedra and the unrotated ones, with respect to $P6_3cm$-BaTiS$_3$, are highlighted in red and blue, respectively, in **b,c**. The bottom panels in **b,c** show a parallel alignment of the Ti off-center displacements along the *c*-axis, with the net polarization point up in **b**, and down in **c**. Barium, titanium and sulfur atoms are shown in orange, blue and grey, respectively.

**Direct observation of ferroelectric and ferroaxial order in BaTiS$_3$**

To verify the theoretical predictions, we synthesized BaTiS$_3$ single crystals using an iodine-assisted vapor transport method, which has been reported earlier[31]. The typical way to impose biaxial strains is through the growth of thin films that are epitaxially strained to a substrate of appropriate dimensions. The growth of epitaxially strained thin films of ternary chalcogenides is still in a nascent state[32-35]. A recent report demonstrated the growth of BaTiS$_3$ thin films that were found to be polycrystalline in nature[34]. To overcome these challenges, we used an innovative strategy involving focused ion beam (FIB) milling of a confined single crystal to introduce >3% biaxial strain. Briefly, this process involves attaching an appropriately oriented single crystal to a TEM post such that the two opposite lateral sides of the single crystal are fixed to provide the confinement, instead of the conventional approach of attaching the specimen to the TEM post with only one side fixed. This was followed by plasma FIB milling to make a thin, electron-transparent lamella (more details in *Supplementary Section 3* Fig. S3). The FIB milling also introduces stresses due to ion impact and surface amorphization[36-37]. For the TEM lamellae with two sides fixed, we expect the intrinsic stress from the FIB milling will be confined to trigger the phase transition, whereas the TEM lamellae with only one side fixed to the post, the residual stresses can relax through the unattached end. We repeated this process three times and could reproducibly achieve a biaxially strained sample with strains ranging between 3.08 – 3.21 %.

To directly resolve the crystal structure of the strained BaTiS$_3$ sample and compare with the unstrained sample, we performed atomically resolved imaging using an aberration-corrected STEM. A high-angle annular dark field (HAADF) STEM image of a typical strain-relaxed lamella of *P*6$_3$*cm*-BaTiS$_3$ viewed along [001] is shown in Fig. 2(b). In this imaging mode, the intensity of the atomic columns is proportional to the square of their effective atomic number ($Z^2$)[38]. The heaviest Ba columns appear brightest and form a regular hexagonal network. The lighter Ti columns have lower intensity, and are present about the centers of the Ba hexagons. A quantitative analysis of the Ba and Ti atomic columns shows that the Ti atoms are off-centered with respect to the centroids of the surrounding Ba hexagons. These Ti displacements are ordered locally over 2-5 nm, but are disordered over the field of view of the image (see Fig. S4 in *Supplementary Section 4*). These observations are in good agreement with a recent study[26]. The lightest S columns around the Ti atoms show up as weak streaks.

A HAADF image of the *ab*-plane of a typical biaxially strained lamellae is shown in Fig. 2(c). In contrast to the corresponding image of *P*6$_3$*cm*-BaTiS$_3$ [Fig 2(b)], the strained structure shows a periodic distortion of the TiS$_6$ octahedra involving a ~30-degree rotation. The Ba atoms also undergo breathing displacements along the *ab*-plane to form irregular hexagons. Together, the TiS$_6$ octahedral rotations and *ab*-plane displacements of Ba result in a mirror-symmetry breaking parallel to the 6-fold rotation axis, as we had discussed for the DFT-optimized *P*6$_3$-BaTiS$_3$ chiral multiferroic phase [Fig. 1(b-c)]. The features of *P*6$_3$-BaTiS$_3$ in the strained crystal were further confirmed by comparing them with the corresponding simulated HAADF image of a DFT-optimized structure, and its FFT analysis, as shown in Fig. S5(c-d, g-h). We find excellent agreement between the simulated and experimental images. To better resolve the rotated TiS$_6$ octahedra in *P*6$_3$-BaTiS$_3$, we performed differential phase contrast (DPC) imaging by collecting four-dimensional (4D) datasets using a pixelated detector[39-41]. The reconstructed atomic potential images (see Fig. S6 in *Supplementary Section 4*) provide a clear view of the lighter sulfur atomic columns forming rotated and unrotated TiS$_6$ octahedra in the strained BaTiS$_3$ sample.

To gain further insights into the local structure and identify the different configurations of the Ba sublattice in the two samples, we used principal component analysis (PCA) (details can be found in *Supplementary Section 5*). The scree plot obtained after performing PCA on the HAADF image of $P6_3cm$-BaTiS$_3$ shows a gradual variation, suggesting similar local configurations with only small variations in the spacing between Ba columns. In contrast, we find four distinct components in the scree plot of the strained $P6_3$-BaTiS$_3$ [Fig. S7(c)] sample, suggesting four different local configurations. We used unsupervised $k$-means clustering analysis to group Ba atoms based on the similarity of the local structural motifs centered around TiS$_6$ columns (6 Ba atoms per structural motif). The four local structural motifs in [001] $P6_3$-BaTiS$_3$ are shown in Figs. 2(d-g). In the strained BaTiS$_3$ sample, we observe nanoscale (~15 nm) domains consisting of three distorted Ba sublattice configurations, with the domains being separated by sharp interfaces consisting of a less distorted Ba sublattice [Fig. S8(d)]. We extracted the atomic displacements in the four motifs by comparing the column positions to that of the regular hexagons in $P6_3cm$-BaTiS$_3$. As shown in Fig. 2(b), we find the Ba atoms in $P6_3$-BaTiS$_3$ undergo different kinds of displacements in the *ab*-plane. For the rotated TiS$_6$ octahedron shown in Fig. 2(f), its neighboring Ba atoms undergo outward displacements, which expand the Ba hexagon. These outward displacements subsequently lead to a dimerization of Ba atoms, as shown in Figs. 2(d-e). The structural motif in Fig. 2(g) was extracted from the sharp domain boundaries, as shown in Fig. S8(d). These local symmetry features of $P6_3$-BaTiS$_3$, as resolved by atomic-resolution STEM characterization, exhibit a consistent match with the DFT-optimized $P6_3$-BaTiS$_3$ lattice, as shown in Fig. S1(d-f).

We used the Ba–Ba distances from the HAADF images of the *ab*-plane of $P6_3cm$-BaTiS$_3$ and $P6_3$-BaTiS$_3$ to measure the biaxial strain in the latter. As opposed to the uniform Ba–Ba distances in $P6_3cm$-BaTiS$_3$ [Fig. 2(h)], there is substantial modulation of Ba–Ba distances in $P6_3$-BaTiS$_3$ [Fig. 2(i)]. The nanoscale domains (of around ~15 nm size) are found to be separated by sharp interfaces consisting of hexagons with shorter Ba–Ba distance [Fig. 2(i)]. A similar domain pattern was also revealed in our $k$-means clustering analysis in Fig. S8(d). Notably, the average Ba–Ba distance in $P6_3$-BaTiS$_3$ is measured to be larger than $P6_3cm$-BaTiS$_3$ by 3.08%, which implies a biaxial tensile strain state in $P6_3$-BaTiS$_3$ [Fig. 2(j)], and is consistent with the DFT-predicted stability of the latter under tensile strains. The strain results were also obtained from Ti–Ti distance mapping for $P6_3cm$-BaTiS$_3$ and $P6_3$-BaTiS$_3$, as shown in Fig. S9, and are consistent with the Ba—Ba distance results. More domain patterns can be seen from a larger field of view observation of $P6_3$-BaTiS$_3$ in Fig. S10(e-f).

Having confirmed the ferroaxial distortions along the *ab*-plane in the strained sample, we next performed STEM imaging of the strained and strain-relaxed samples oriented along [1$\bar{1}$0] to probe the predicted change from antipolar Ti displacements along the *c*-axis in $P6_3cm$-BaTiS$_3$ to polar displacements in $P6_3$-BaTiS$_3$. The antipolar displacements in $P6_3cm$-BaTiS$_3$ lead to a separation of the S and Ti columns along the *c*-axis, as shown in Fig. 3(a). As a result, the horizontal intensity profile between two adjacent Ba atomic columns exhibits only one peak in the middle corresponding to an S atomic column [Fig.3(c, e)], with the Ti atomic column displaced either upwards or downwards. In the HAADF image of the strained $P6_3$-BaTiS$_3$ lattice shown in Fig. 3(b), the polar displacements of Ti atoms along the *c*-axis aligns them with the S atomic columns (the Ti and S columns are offset along the *a*-axis). Consequently, two peaks, instead of one, show up in a horizontal line profile of the HAADF intensity across two Ba atoms [Fig. 3(d, e)]. The experimentally observed changes in intensity are in excellent agreement with the simulated STEM results for $P6_3cm$-BaTiS$_3$ and $P6_3$-BaTiS$_3$ oriented along [1$\bar{1}$0], as shown in Fig. 3(f). Overall, using STEM, we are able to confirm the presence of both ferroaxial and ferroelectric order in the strained sample of $P6_3$-BaTiS$_3$.

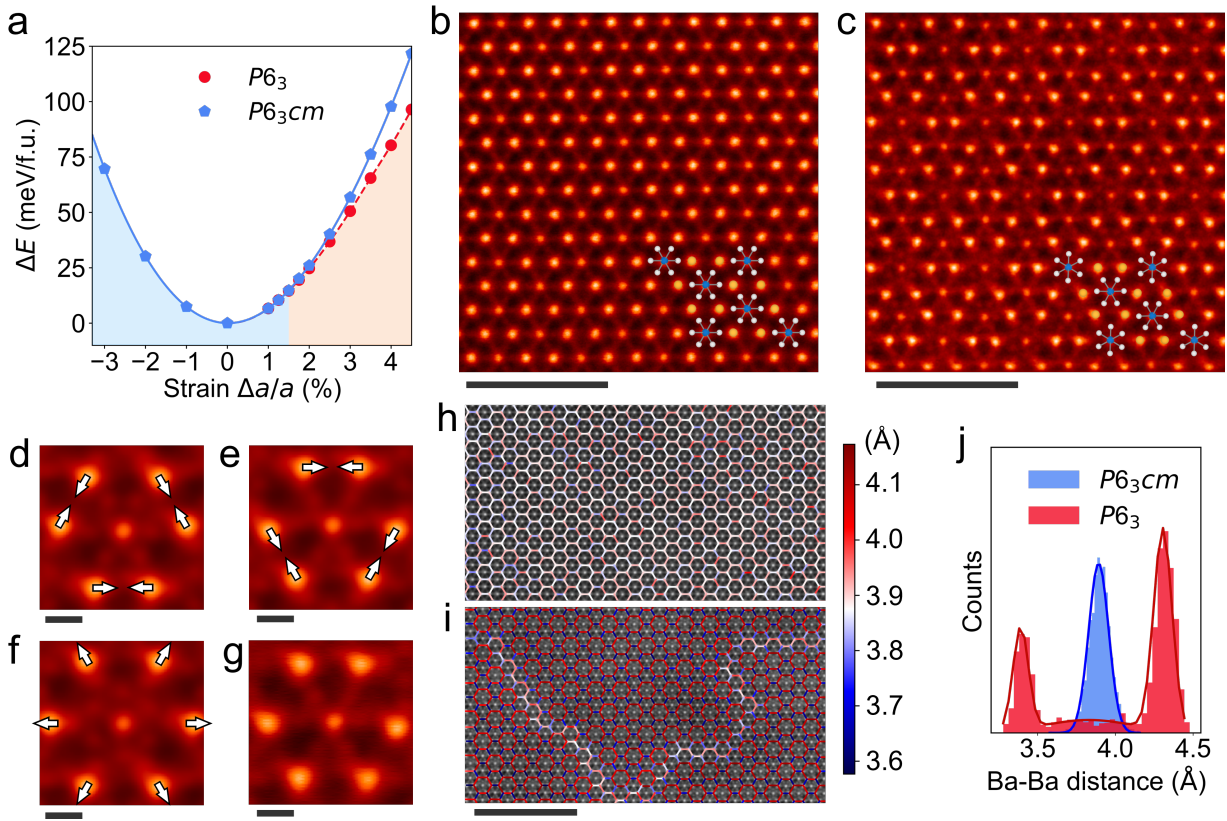

**Figure 2. Strain-induced stabilization and atomic-scale observation of ferroaxial order in BaTiS$_3$. a.** Phase stability as a function of biaxial strain on the *ab*-plane. *P*6$_3$ is stable over *P*6$_3$*cm* for tensile strains exceeding 1.5%. A representative atomic-resolution STEM-HAADF image of *P*6$_3$*cm*-BaTiS$_3$ in **b** and biaxially strained *P*6$_3$-BaTiS$_3$ in **c**. Both the samples are viewed along the [001] axis. The structure models of *P*6$_3$*cm* and *P*6$_3$-BaTiS$_3$ phases are overlaid onto the STEM-HAADF images to provide clearer identification of the atomic columns. **d-g.** The top four principal components extracted from a *k*-means clustering analysis of the HAADF image of *P*6$_3$-BaTiS$_3$. The overlaid white arrows represent the displacement of Ba atomic columns relative to the undistorted hexagonal Ba sublattice in *P*6$_3$*cm*-BaTiS$_3$. The rotational distortion of TiS$_6$ octahedron can be seen in **f**, and contrasted with unrotated TiS$_6$ octahedra in **d, e,** and **g**. Ba–Ba distance mapping for *P*6$_3$*cm*-BaTiS$_3$ in **h** and *P*6$_3$-BaTiS$_3$ in **i**. The overlaid lines connect neighboring Ba atoms. The magnitude of Ba–Ba distance is represented using the color-scale shown on the right. **j.** Histogram plot showing the distribution of Ba–Ba distances in *P*6$_3$*cm*-BaTiS$_3$ and *P*6$_3$-BaTiS$_3$. *P*6$_3$*cm*-BaTiS$_3$ shows a unimodal distribution with an average Ba–Ba distance of 3.9 Å, while *P*6$_3$-BaTiS$_3$ has a bimodal distribution with an average value of 4.02 Å. It is equivalent to an average biaxial tensile strain of 3.08 %. The histograms were fitted with Gaussian curves shown with solid lines. Scale bars are 2 nm in (**b-c**), 2 Å in (**d-g**), and 5 nm in (**h-i**).

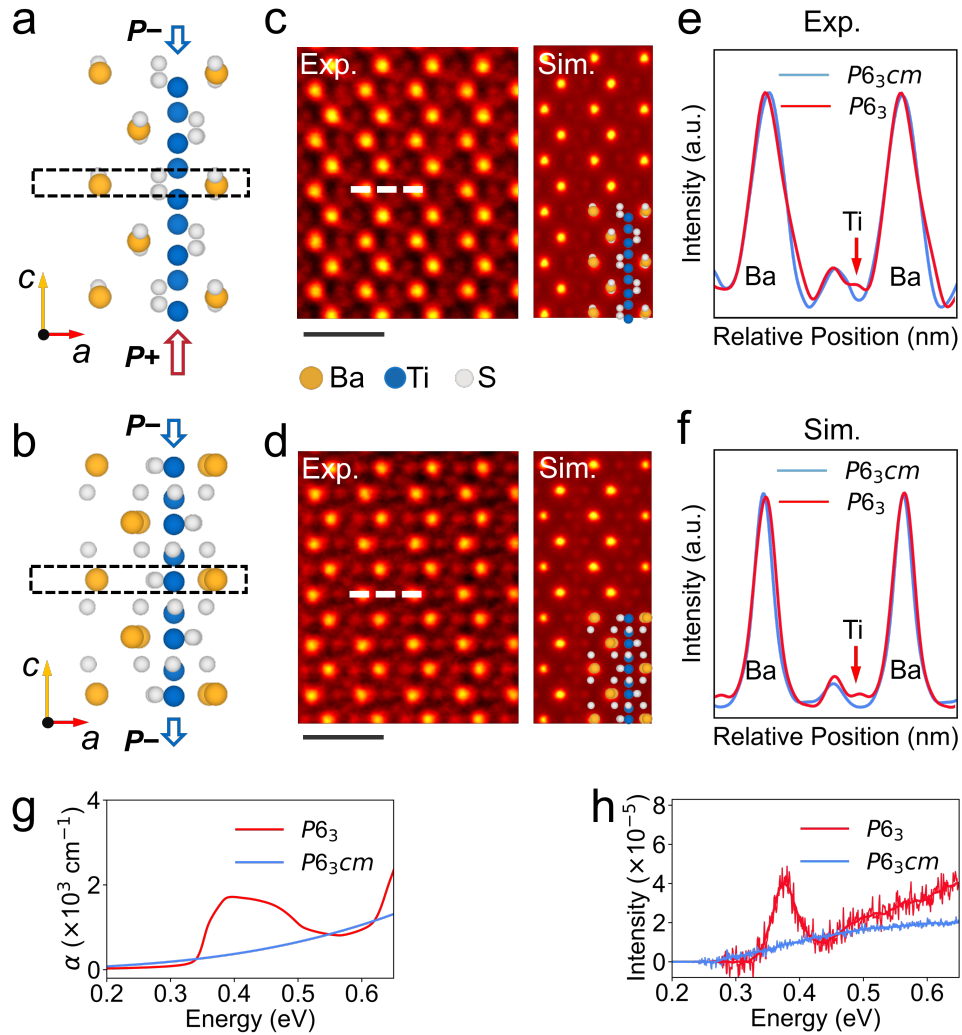

**Figure 3. Atomic-scale structures of BaTiS$_3$ viewed along the [1$\bar{1}$0] zone axis and valence EELS showing changes in electronic structure with the phase transition. a.** Atomic arrangement in $P6_3cm$-BaTiS$_3$ along the $c$-axis, where S atoms and Ti atoms misalign with each other along the horizontal direction within the dashed box region. **b.** Atomic arrangement in $P6_3$-BaTiS$_3$ having ferroelectric distortion of Ti atoms along the $c$-axis, which aligns the S atoms with Ti along the horizontal direction shown with the dashed box. **c-d.** Experimental and simulated STEM-HAADF images of $P6_3cm$-BaTiS$_3$ and $P6_3$-BaTiS$_3$, both oriented along the [1$\bar{1}$0] zone axis showing the atomic arrangements between the Ba columns. **e-f.** Horizontal intensity profiles along the dashed lines in the experimental and simulated STEM images of $P6_3cm$-BaTiS$_3$ and $P6_3$-BaTiS$_3$ lattices in **c** and **d**, respectively. Compared to the single peak due to S atomic columns between Ba peaks in $P6_3cm$-BaTiS$_3$, an additional peak corresponding to Ti atomic columns indicated by the red arrow shows up in $P6_3$-BaTiS$_3$. **g.** DFT calculated absorption spectra of $P6_3cm$-BaTiS3 and $P6_3$-BaTiS$_3$ for comparison. **h.** Monochromated valence EELS measurements for valence electron excitation showing the strong absorption peak in $P6_3$-BaTiS$_3$ which is consistent with DFT calculations in **g**. Scale bars are 1 nm in (**c-d**).

While the STEM-HAADF images show the presence of multiple domains in the strained $P6_3$-BaTiS$_3$ sample [see Fig. 2(i)], it does not provide evidence that the different domains have different chirality. Chiral domains can be mapped by scanning a circularly polarized light over the sample, which will interact differently with domains of different handedness or chirality[13-15, 18]. However, the 15 – 20 nm wide domains in strained $P6_3$-BaTiS$_3$ are much smaller than the spot size of a typical laser, thus making it challenging to map their chirality using polarized light. Instead, we used convergent beam electron diffraction (CBED) in a STEM to map domains having different chirality. STEM-CBED is a powerful tool to extract information about the local crystal structure and symmetry[42-44]. We used a nanometer-sized (~1 nm) electron probe and rastered it across the $P6_3$-BaTiS$_3$ lamella (see experimental details in *Methods*). At each probe position (2 dimensions in the real space), we recorded the diffraction pattern (2 dimensions in the reciprocal space) using a pixelated direct-electron detector. The acquired 4D dataset, formed by zero-order Laue zone patterns, includes information about the rotational and mirror symmetries of the structure[43-44], and was used to extract the distribution of chiral domains. We carried out unsupervised *k*-means clustering analysis to automatically separate the domains based on the symmetry information recorded in each CBED pattern. As shown in Fig. S11(a-b), two distinctive regions possessing different rotational symmetries are identified in the scanned area. The corresponding CBED patterns from these two regions, as shown in Fig. S11(c-d), demonstrate distinct 3-fold rotational symmetry originating from $6_3$ screws in $P6_3$-BaTiS$_3$. Moreover, these two CBED patterns are related through a mirror reflection with the mirror plane being parallel to (110). This mirror operation is one of the symmetry elements which is present in the high-symmetry $P6_3cm$-BaTiS$_3$ phase, but is lost in the $P6_3$-BaTiS$_3$ phase due to the ferroaxial distortions involving clockwise or counterclockwise rotation of TiS$_6$ octahedra. Thus, the STEM-CBED experiments provide confirmation that the domains identified using STEM-HAADF imaging indeed have different chirality. We also performed composition analysis of the two samples using atomically resolved electron energy-loss spectroscopy (EELS) analyses (see details in *Supplementary Section 7* Fig. S12), and observed no elemental segregation in either sample, even at the domain boundaries in the chiral multiferroic phase. It confirms that the observed ferroelectric ferroaxial order in $P6_3$-BaTiS$_3$ is a displacive-type transition, as opposed to the order-disorder type transition reported in NiTiO$_3$, a ferroaxial material[15].

**Electronic structure of chiral multiferroic $P6_3$-BaTiS$_3$**

Having successfully demonstrated the presence of ferroelectric ferroaxial distortions that lead to chiral multiferroic ordering in biaxially strained $P6_3$-BaTiS$_3$, we next discuss its electronic properties (see details in *Supplementary Section 8*). We observe a slight increase in the theoretical bandgap of $P6_3$-BaTiS$_3$ (0.32 eV) with 3.5 % strain over unstrained $P6_3cm$-BaTiS$_3$ (0.24 eV), calculated at the PBE level (see Fig. S13). The ferroaxial distortions in $P6_3$-BaTiS$_3$ introduce localized S-3$p$ states originating from rotated TiS$_6$ octahedra [Fig. S13(d)], which results in flat valence bands close to Fermi energy [Fig. S13(c)]. To confirm the presence of the flat bands, we performed valence EELS measurements on the two samples using a monochromated STEM. The normalized EELS results for valence electron excitations are compared in Fig. 3(h). We observe a prominent absorption peak at ~0.34 eV in $P6_3$-BaTiS$_3$, due to the flat bands, which are absent in $P6_3cm$-BaTiS$_3$. The corresponding EELS spectra exhibit excellent agreement with the absorption spectra of band edge excitations calculated using DFT in Fig. 3(g). The deviations between the EEL and the DFT-calculated spectra at higher energies (above 0.45 eV) may be attributed to the coexistence of other energy excitations, such as collective excitations of valence electrons (plasmons) and Cerenkov radiation[45], and the instrumental broadening, all of which tend to smear the spectroscopic features. Additionally, the influence the chiral domain boundaries can also affect the absorption features in strained $P6_3$-BaTiS$_3$, which warrants further investigation.

**Coupling between ferroelectric and ferroaxial order in $P6_3$-BaTiS$_3$**

We now discuss the origin of the ferroelectric ferroaxial transition and the coupling between the two order parameters in strained $P6_3$-BaTiS$_3$. As mentioned before, each of the order parameters has two degenerate configurations (*P+/P–* and *A+/A–*), and together they lead to four domains (*A+P+* to *A−P−*, or *A+P−* to *A−P+*) that are degenerate in energy. To estimate the barriers related to the transition between the four configurations, we combined DFT calculations with a Landau model of phase transitions[46]. The free energy of strained BaTiS$_3$, *F*, was expanded as a power series of the two primary order parameters, *A* and *P* (see details in *supplementary section 9*). We used the centrosymmetric and achiral $P6_3/mmc$ structure as the reference phase for the transition between opposite ferroelectric ferroaxial order (*A+P+* to *A−P−*, or *A+P−* to *A−P+*). Symmetry analysis confirms that the transition from $P6_3/mmc$ to $P6_3$ is a valid ferroelectric ferroaxial transition[19]. We note that the actual switching pathway in the experiments may involve complex domain boundaries or other metastable phases[25, 29], and using the $P6_3/mmc$ structure as the saddle point likely gives an upper bound of the switching barriers. The amplitudes of ferroelectric and ferroaxial modes, $Q_P$ and $Q_A$, have been normalized to those in $P6_3$-BaTiS$_3$ under 3.5% tensile strain, which is close to the experimentally measured strain. The free energy expression for the coupled transition in terms of $Q_P$ and $Q_A$ up to their eighth order coupling can then be written as follows:

$$F(Q_P, Q_P) = (\alpha_A Q_A^2 + \beta_A Q_A^4 + \gamma_A Q_A^6) + (\alpha_P Q_P^2 + \beta_P Q_P^4 + \gamma_P Q_P^6) \\ + (\delta_{A \cdot P} Q_A^2 Q_P^2 + \kappa_{A \cdot P} Q_A^4 Q_P^4),$$

(Eq.1)

where ($\alpha_A$, $\beta_A$, $\gamma_A$) and ($\alpha_P$, $\beta_P$, and $\gamma_P$) are, respectively, the coefficients for individual ferroaxial and ferroelectric mode, while $\delta_{A-P}$ and $\kappa_{A-P}$ are coefficients accounting for the coupling between the two modes. Fig. 4a shows the DFT-calculated free energy as a function of the amplitude of the individual ferroelectric mode ($Q_P$), ferroaxial mode ($Q_A$), and the coupled mode in strained $P6_3$-BaTiS$_3$. We find that ferroelectric $Q_P$ is characterized by a typical double-well potential, and by itself, leads to a relatively modest lowering of free energy over the reference strained $P6_3/mmc$-BaTiS$_3$ phase. In contrast, the ferroaxial mode, $Q_A$, by itself, is unstable as it forms a single-well potential with a minimum at $Q_A = 0$, and increasing energy for finite $Q_A$. However, when the two modes are coupled, we observe an appreciable lowering of energy for $P6_3$-BaTiS$_3$, which indicates a strong coupling between the two order parameters. Energy contributions for the $P6_3$-BaTiS$_3$ transition were calculated to be 658.0 meV/f.u. for pure $Q_P$, 311.6 meV/f.u. for pure $Q_A$, and -1066.3 meV/f.u. for the coupling term, respectively (Fig. S17; The fitted Landau coefficients can be found in Table S1 in *supplementary section 9*). It also implies that the chirality is improper in nature, as it is stabilized only due to its strong coupling with the ferroelectric mode. The resulting energy landscape between the four degenerate chiral multiferroic states, which is obtained using the fitted energy model, is shown in Fig. 4b. We find that the lowest energy pathway to switch the direction of the polarization $Q_P$ is also expected to switch the handedness or the chirality $Q_A$, due to the strong coupling between the two modes. Thus, we predict that $P6_3$-BaTiS$_3$ should allow for electric-field tunability of chirality.

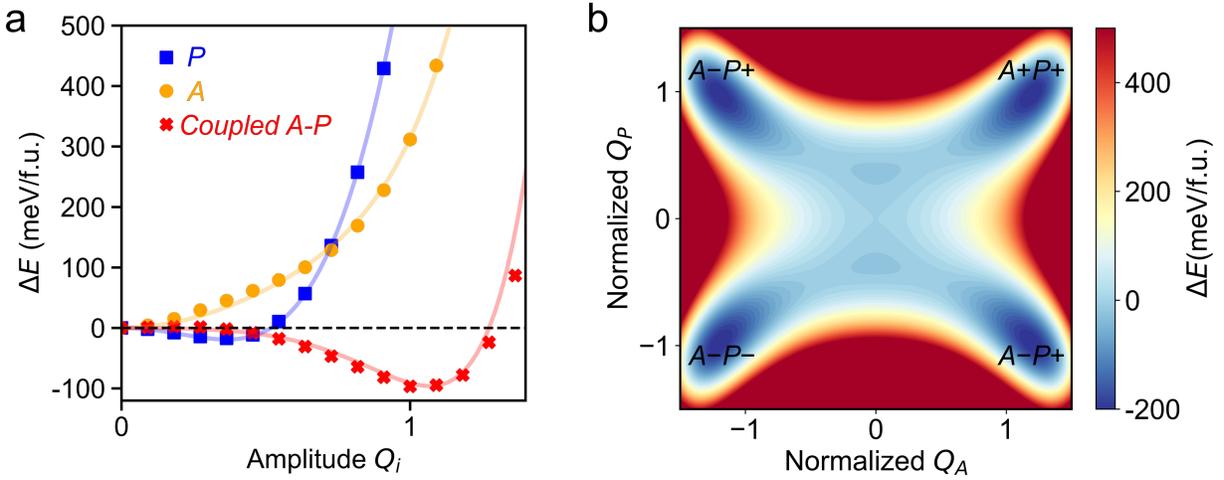

**Figure 4. Landau free energy landscape for the coupled ferroelectric ferroaxial transition. a.** DFT-calculated potential energy of the primary modes ($Q_P$ and $Q_A$), and the coupled distortion mode in strained $P6_3$-BaTiS$_3$. The normalized distortion amplitude ($Q_i$) = 0 corresponds to the centrosymmetric $P6_3/mmc$ phase, while 1 corresponds to the $P6_3$ phase, with both the phases under 3.5% biaxial strain. **b**. Contour plot of the Landau free energy as a function of the primary order parameters ($Q_A$, $Q_P$). The four degenerate chiral multiferroic states are labeled.

**Conclusion and outlook**

In summary, we have demonstrated the successful stabilization of a room-temperature chiral multiferroic phase in single-crystals of quasi-1D $P6_3$-BaTiS$_3$ under biaxial tensile strain. This multiferroic state includes ferroelectric order along the long *c*-axis, and ferroaxial order around the same axis, both of which were confirmed using atomic-resolution STEM imaging, diffraction and EELS. We observe the presence of nanoscale chiral domains that are separated by sharp domain boundaries. In pure ferroaxial materials, it is non-trivial to switch their chirality as the ferroaxial order is invariant under both time-reversal and space-inversion operations, and hence, does not couple to typical thermodynamic fields. However, in $P6_3$-BaTiS$_3$, we predict the presence of a strong coupling between the ferroelectric and ferroaxial order. In fact, the ferroaxial order, by itself, is found to be unstable, and is stabilized due to the coupling with the ferroelectric order in $P6_3$-BaTiS$_3$. Thus, we predict $P6_3$-BaTiS$_3$ to be a strong candidate to show electric-field tuning of chirality.

To demonstrate electric-field control of chirality in $P6_3$-BaTiS$_3$, several open questions will need to be addressed. Our STEM studies show that the chiral domains in strained $P6_3$-BaTiS$_3$ are extremely small, spanning about 15 – 20 nm in size. Such small domains pose a challenge for probing using conventional optical probes. It might be possible to probe their chiral properties using electron energy-loss magnetic chiral dichroism carried out within a STEM (STEM-EMCD)[47], or through electron vortex beams[48], which are sensitive to local orbital angular momentum. Annealing the material might lead to a reduction in the chiral domain walls and increase the size of the chiral domains making them suitable for probing using conventional optical techniques. Lastly, many ternary oxides and chalcogenides adapt the hexagonal perovskite structure. It might be possible to stabilize them in a chiral multiferroic phase using strain to obtain electric-field tunable chiral materials for different wavelengths and applications. In fact, the oxide analogue of BaTiS$_3$, i.e., hexagonal BaTiO$_3$ already undergoes a ferroelectric ferroaxial transition below 74 K[22]. It might be possible to push the transition temperature of BaTiO$_3$ towards room temperature through epitaxial growth, which is well established for oxides.

**Methods**

**Sample preparation**

Single crystals of BaTiS$_3$ were synthesized using an iodine-assisted chemical vapor transport (CVT) method as reported earlier[25-28, 31]. [001]-orientated TEM lamellae were prepared using a Thermo Scientific Helios G4 Plasma focused ion beam (PFIB) UXe Dual Beam system. A Pt layer was deposited over the region of interest prior to ion milling. Xe PFIB was used to make TEM sample with a final thickness of ~100 nm using 30 kV and 0.1 nA as the final voltage/current. TEM lamella with different strain states were prepared through a standard lift-out routine: one [001]-BaTiS$_3$ lamella was welded onto a TEM half grid with one side fixed such that strain relaxation was possible through the unconstrained side. For another lamella, we attached it to a TEM half-grid with two lateral sides fixed such that biaxial strain can be introduced inside the lamella upon thinning. All the TEM lamellae were further thinned down to an electron transparent thickness utilizing a Fischione's Model 1040 NanoMill TEM specimen preparation system at an operating voltage of 900 V and at incident angels of 5° and -5°.

**Electron microscopy**

Atomic resolution *Z*-contrast STEM-HAADF imaging was carried out using a Nion UltraSTEM™ equipped with a spherical aberration corrector operating at 100 kV. The STEM-HAADF images were acquired with a convergence semi-angle of 30 mrad and collection semi-angles from 80 to 200 mrad. To correct the sample drift effect, we recorded each STEM dataset as a series of 20 successive frames with fast scanning (1 μs/pixel dwell time) for post-processing correction through image registration. STEM-HAADF simulations were performed using the multi-slice method as implemented in μSTEM[49]. *P*6$_3$*cm* and *P*6$_3$ crystal structures optimized by DFT were used for the STEM simulations. To match the experimental conditions, we performed the simulations using an aberration-free probe with an accelerating voltage of 100 kV and a convergence semiangle of 30 mrad. Thermal scattering was included in our simulation through the phonon excitation model proposed by Forbes et al[50]. The sample thickness was set to 15 nm and the defocus value was set to 10 Å to obtain good agreement with the experimental results.

Lattice distortions were obtained from the atomic-scale STEM-HAADF images by mapping the atomic distances between neighboring Ti—Ti and Ba—Ba atoms. Atomic positions were determined by fitting 2D Gaussian peaks to Ti and Ba columns. The local crystallography for Ba and Ti sublattices was mapped out using data analytics tools available within the Pycroscopy suite[51]. All the image data analyses and visualizations were carried out using the home-developed python-based codes.

4D-STEM experiments were carried out on a Nion UltraSTEM™ at 100 kV using a Dectris ELA high-speed electron-counting detector. Differential phase contrast (STEM-DPC) datasets were acquired with 256 × 256 pixel size using a 30 mrad convergent beam. At each pixel position, a 64 × 64 pixel Ronchigram image was recorded with a dwell time of 0.5 ms. Convergent-beam electron diffraction (STEM-CBED) patterns were obtained with a probe semi-angle of 2.5 mrad to avoid the overlap of Bragg discs. Diffraction patterns were collected using a step size of 1 nm with 32 × 32 scan positions. Dwell time for each pixel was set at 2 ms. Each diffraction pattern was recorded on Dectris ELA detector with frame size of 512 × 512 pixels. $k$-means clustering was applied for differentiating the local rotational and mirror symmetries of the projected structure.

Core-loss electron energy loss spectra (EELS) were acquired with pixel dwell times of 3 ms, energy dispersion of 1 eV per channel, and a collection semi-angle of 35 mrad on a Nion Iris spectrometer attached to the Nion UltraSTEM. Ba-$M_{4,5}$, Ti-$L_{2,3}$, and S-$L_{2,3}$ edges were used for elemental quantification. The spectrum images were denoised using principal component analysis (PCA). A power law was used to model the background signal prior to the core-loss signal. Core-loss EELS data were processed using Digital Micrograph.

Electronic structures of $P6_3cm$-BaTiS$_3$ and $P6_3$-BaTiS$_3$ were characterized by valence EELS acquired at an operating voltage of 100 kV using a Nion IRIS EEL spectrometer attached to the Nion aberration-corrected monochromated STEM-EELS (Ultra-HERMES™) at Oak Ridge National Laboratory (ORNL). The Ultra-HERMES™ at ORNL is equipped with a Dectris ELA high-speed electron-counting detector. The probe convergence semi-angle of 32 mrad, and an entrance aperture collection angle of 25 mrad were used to record the spectra. An energy dispersion of 0.85 meV per channel was selected to achieve an energy resolution of 13.2 meV and 18.7 meV for unstrained and strained BaTiS$_3$ samples, respectively (see Supplementary Fig. S15). The spectrum-images (512 × 512 pixels) were recorded with dwell times of 10 ms per pixel. To process the valence EELS spectra, we first calibrated the recorded EEL spectra by aligning the center of the zero-loss peak (ZLP) at 0 meV using a correlation-based routine as implemented in the Nion Swift package. For a good approximation of scattering probability, all the EELS signals were normalized to the ZLP intensity. Background subtraction was performed by fitting a power-law function with a fitting window from 0.25 to 0.30 eV in low-loss EELS spectra.

**Density-functional theory calculations**

First-principles calculations based on density functional theory were performed using the Vienna Ab initio Simulation Package (VASP)[52]. We used projector augmented-wave (PAW) potentials to treat the electron-ion interactions. The generalized gradient approximation within the Perdew-Burke-Ernzerhof (GGA-PBE) functional was adopted for exchange-correlation interactions[53-55]. A plane-wave basis set with a cutoff energy of 650 eV was used according to convergence tests. The Brillouin zone was sampled using a $\Gamma$-centered $k$-point mesh with spacing of 0.025 Å$^{-1}$. Convergence for electronic calculations was reached with energy change under $10^{-7}$ eV. To investigate the effect of strain on the coupled ferroelectric ferroaxial order, we firstly initialized rotational distortion of TiS$_6$ octahedra in strained $P6_3$-BaTiS$_3$ lattice and then optimized the cell volume along the $c$ axis under biaxial strain conditions. The ionic positions were relaxed until the forces were smaller than $10^{-3}$ eV/Å.

To compare the valence EELS results, the frequency-dependent dielectric function $\varepsilon(\omega)=\varepsilon_1(\omega)+i\varepsilon_2(\omega)$ of $P6_3$-BaTiS$_3$ and unstrained $P6_3cm$-BaTiS$_3$ crystals were simulated using the independent particle approximation (IPA)[56]. The imaginary part $\varepsilon_2(\omega)$ is obtained by calculating the direct transitions between occupied and unoccupied states. The absorption coefficient α is calculated as $\omega\varepsilon_2/(nc)$ (cm$^{-1}$, Gaussian unit), where $n = \sqrt{(\sqrt{\varepsilon_1^2 + \varepsilon_2^2} + \varepsilon_1)/2}$ is the refractive index and $c$ is the speed of light in vacuum. To match the energy resolution with valence EELS spectra, we set the CSHIFT tag in VASP to 0.01 and NEDOS = 10000.

The distortion modes of $P6_3$-BaTiS$_3$ from the centrosymmetric $P6_3/mmc$-BaTiS$_3$ structure were investigated using the ISODISTORT Software Suite[57-58].

**Acknowledgements**

This work was supported, in part, by an ARO MURI program with award no. W911NF-21-1-0327, and the National Science Foundation (NSF) of the United States under grant numbers DMR-2122070 (G.Y.J.), DMR-2122071, and DMR-2145797 (R.M.). M.C., A. R. L., and J. A. H. are supported by the U.S. Department of Energy, Office of Science, Basic Energy Sciences, Materials Sciences and Engineering Division. The Microscopy work was conducted as part of a user project at the Center for Nanophase Materials Sciences (CNMS), which is a DOE Office of Science User Facility using instrumentation within ORNL's Materials Characterization Core provided by UT-Battelle, LLC, under Contract No. DE-AC05-00OR22725 with the DOE and sponsored by the Laboratory Directed Research and Development Program of Oak Ridge National Laboratory, managed by UT-Battelle, LLC, for the U.S. Department of Energy. This work used computational resources through allocation DMR160007 from the Advanced Cyberinfrastructure Coordination Ecosystem: Services & Support (ACCESS) program, which is supported by NSF grants #2138259, #2138286, #2138307, #2137603, and #2138296.

**Author Contributions**

G.R. and R.M. conceived the idea and designed the experiments. H.C., B.Z. and J.R. synthesized the single crystals of BaTiS$_3$. H.C. performed PFIB sample preparations. G.R. and R.M. carried out all the STEM measurements and their analyses. R.K.V., J.A.H., A.R.L., and M.C. supervised STEM experiments and analyses. G.Y.J., G.R., and R.M. performed the first-principles DFT calculations and Landau free energy analyses. C.W. and D.X. contributed to the discussion of theoretical results. G.R. and R.M. drafted the manuscript with edits from all authors.

**Competing interests**

The authors declare no conflict of interest.

**Data availability**

Data presented in the main text of this paper are openly available in Zenodo at https://doi.org/10.5281/zenodo.12810613. Data presented in the Supplementary Information are available upon request from the corresponding authors.

**References**

[1] N. A. Spaldin, R. Ramesh, *Nature materials* **2019**, 18, 203.

[2] M. Coll, J. Fontcuberta, M. Althammer, M. Bibes, H. Boschker, A. Calleja, G. Cheng, M. Cuoco, R. Dittmann, B. Dkhil, *Applied surface science* **2019**, 482, 1.

[3] C. Sang-Wook, T. Diyar, K. Valery, S. Avadh, *NPJ Quantum Materials* **2018**, 3.

[4] G. Chang, B. J. Wieder, F. Schindler, D. S. Sanchez, I. Belopolski, S.-M. Huang, B. Singh, D. Wu, T.-R. Chang, T. Neupert, *Nature materials* **2018**, 17, 978.

[5] J. Junquera, Y. Nahas, S. Prokhorenko, L. Bellaiche, J. Íñiguez, D. G. Schlom, L.-Q. Chen, S. Salahuddin, D. A. Muller, L. W. Martin, *Reviews of Modern Physics* **2023**, 95, 025001.

[6] A. Fert, N. Reyren, V. Cros, *Nature Reviews Materials* **2017**, 2, 1.

[7] V. Gopalan, D. B. Litvin, *Nature materials* **2011**, 10, 376.

[8] W. Jin, E. Drueke, S. Li, A. Admasu, R. Owen, M. Day, K. Sun, S.-W. Cheong, L. Zhao, *Nature Physics* **2020**, 16, 42.

[9] J. Hlinka, *Physical Review Letters* **2014**, 113, 165502.

[10] B. B. Van Aken, J.-P. Rivera, H. Schmid, M. Fiebig, *Nature* **2007**, 449, 702.

[11] R. Johnson, L. Chapon, D. Khalyavin, P. Manuel, P. Radaelli, C. Martin, *Physical review letters* **2012**, 108, 067201.

[12] R. Johnson, S. Nair, L. Chapon, A. Bombardi, C. Vecchini, D. Prabhakaran, A. Boothroyd, P. Radaelli, *Physical review letters* **2011**, 107, 137205.

[13] T. Hayashida, K. Kimura, T. Kimura, *Proceedings of the National Academy of Sciences* **2023**, 120, e2303251120.

[14] T. Hayashida, K. Kimura, D. Urushihara, T. Asaka, T. Kimura, *Journal of the American Chemical Society* **2021**, 143, 3638.

[15] T. Hayashida, Y. Uemura, K. Kimura, S. Matsuoka, D. Morikawa, S. Hirose, K. Tsuda, T. Hasegawa, T. Kimura, *Nature Communications* **2020**, 11, 4582.

[16] D. D. Khalyavin, R. D. Johnson, F. Orlandi, P. G. Radaelli, P. Manuel, A. A. Belik, *Science* **2020**, 369, 680.

[17] A. Yadav, C. Nelson, S. Hsu, Z. Hong, J. Clarkson, C. Schlepütz, A. Damodaran, P. Shafer, E. Arenholz, L. Dedon, *Nature* **2016**, 530, 198.

[18]     P. Behera, M. A. May, F. Gómez-Ortiz, S. Susarla, S. Das, C. T. Nelson, L. Caretta, S.-L. Hsu, M. R. McCarter, B. H. Savitzky, *Science advances* **2022**, 8, eabj8030.

[19]     J. Hlinka, J. Privratska, P. Ondrejkovic, V. Janovec, *Physical review letters* **2016**, 116, 177602.

[20]     D. Di Sante, P. Barone, R. Bertacco, S. Picozzi, *Advanced materials (Deerfield Beach, Fla.)* **2012**, 25, 509.

[21]     F. De Juan, A. G. Grushin, T. Morimoto, J. E. Moore, *Nature communications* **2017**, 8, 15995.

[22]     Y. Akishige, G. Oomi, T. Yamaoto, E. Sawaguchi, *Journal of the Physical Society of Japan* **1989**, 58, 930.

[23]     A. J. Hearmon, F. Fabrizi, L. C. Chapon, R. Johnson, D. Prabhakaran, S. V. Streltsov, P. Brown, P. G. Radaelli, *Physical review letters* **2012**, 108, 237201.

[24]     G. L. Rikken, N. Avarvari, *Nature Communications* **2022**, 13, 3564.

[25]     H. Chen, B. Zhao, J. Mutch, G. Y. Jung, G. Ren, S. Shabani, E. Seewald, S. Niu, J. Wu, N. Wang, *Advanced Materials* **2023**, 35, 2303283.

[26]     B. Zhao, G. Ren, H. Mei, V. C. Wu, S. Singh, G. Y. Jung, H. Chen, R. Giovine, S. Niu, A. S. Thind, *Advanced Materials* **2024**, 2311559.

[27]     S. Niu, G. Joe, H. Zhao, Y. Zhou, T. Orvis, H. Huyan, J. Salman, K. Mahalingam, B. Urwin, J. Wu, *Nature Photonics* **2018**, 12, 392.

[28]     B. Sun, S. Niu, R. P. Hermann, J. Moon, N. Shulumba, K. Page, B. Zhao, A. S. Thind, K. Mahalingam, J. Milam-Guerrero, *Nature Communications* **2020**, 11, 6039.

[29]     B. Zhao, G. Y. Jung, H. Chen, S. Singh, Z. Du, C. Wu, G. Ren, Q. Zhao, N. S. Settineri, S. J. Teat, *arXiv preprint arXiv:2406.09530* **2024**.

[30]     G. H. Fecher, J. Kübler, C. Felser, *Materials* **2022**, 15, 5812.

[31]     B. Zhao, M. S. B. Hoque, G. Y. Jung, H. Mei, S. Singh, G. Ren, M. Milich, Q. Zhao, N. Wang, H. Chen, *Chemistry of Materials* **2022**, 34, 5680.

[32]     M. Surendran, S. Singh, H. Chen, C. Wu, A. Avishai, Y. T. Shao, J. Ravichandran, *Advanced Materials* **2024**, 2312620.

[33]     I. Sadeghi, K. Ye, M. Xu, Y. Li, J. M. LeBeau, R. Jaramillo, *Advanced Functional Materials* **2021**, 31, 2105563.

[34]     M. Surendran, B. Zhao, G. Ren, S. Singh, A. Avishai, H. Chen, J.-K. Han, M. Kawasaki, R. Mishra, J. Ravichandran, *Journal of Materials Research* **2022**, 37, 3481.

[35]     M. Surendran, H. Chen, B. Zhao, A. S. Thind, S. Singh, T. Orvis, H. Zhao, J.-K. Han, H. Htoon, M. Kawasaki, *Chemistry of Materials* **2021**, 33, 7457.

[36]  W. van Mierlo, D. Geiger, A. Robins, M. Stumpf, M. L. Ray, P. Fischione, U. Kaiser, *Ultramicroscopy* **2014**, 147, 149.

[37]  L. Pastewka, R. Salzer, A. Graff, F. Altmann, M. Moseler, *Nuclear Instruments and Methods in Physics Research Section B: Beam Interactions with Materials and Atoms* **2009**, 267, 3072.

[38]  S. Pennycook, D. Jesson, *Ultramicroscopy* **1991**, 37, 14.

[39]  J. A. Hachtel, J. C. Idrobo, M. Chi, *Advanced structural and chemical imaging* **2018**, 4, 1.

[40]  J. Hwang, J. Y. Zhang, J. Son, S. Stemmer, *Applied Physics Letters* **2012**, 100.

[41]  N. Shibata, S. D. Findlay, Y. Kohno, H. Sawada, Y. Kondo, Y. Ikuhara, *Nature Physics* **2012**, 8, 611.

[42]  J. M. LeBeau, S. D. Findlay, L. J. Allen, S. Stemmer, *Ultramicroscopy* **2010**, 110, 118.

[43]  H. Kim, M. Goyal, S. Salmani-Rezaie, T. Schumann, T. N. Pardue, J.-M. Zuo, S. Stemmer, *Physical Review Materials* **2019**, 3, 084202.

[44]  B. Buxton, J. A. Eades, J. W. Steeds, G. Rackham, *Philosophical Transactions of the Royal Society of London. Series A, Mathematical and Physical Sciences* **1976**, 281, 171.

[45]  R. F. Egerton, *Electron energy-loss spectroscopy in the electron microscope*, Springer Science & Business Media, **2011**.

[46]  P. Chandra, P. B. Littlewood, in *Physics of Ferroelectrics: A Modern Perspective*, Springer **2007**, p. 69.

[47]  P. Schattschneider, S. Rubino, C. Hébert, J. Rusz, J. Kuneš, P. Novák, E. Carlino, M. Fabrizioli, G. Panaccione, G. Rossi, *Nature* **2006**, 441, 486.

[48]  J. Verbeeck, H. Tian, P. Schattschneider, *Nature* **2010**, 467, 301.

[49]  L. J. Allen, S. Findlay, *Ultramicroscopy* **2015**, 151, 11.

[50]  B. Forbes, A. Martin, S. D. Findlay, A. J. D'Alfonso, L. J. Allen, *Physical Review B—Condensed Matter and Materials Physics* **2010**, 82, 104103.

[51]  S. Somnath, C. R. Smith, N. Laanait, R. K. Vasudevan, S. Jesse, *Microscopy and Microanalysis* **2019**, 25, 220.

[52]  G. Kresse, J. Furthmüller, *Physical review B* **1996**, 54, 11169.

[53]  J. P. Perdew, K. Burke, M. Ernzerhof, *Physical review letters* **1996**, 77, 3865.

[54]  P. E. Blöchl, *Physical review B* **1994**, 50, 17953.

[55]  G. Kresse, D. Joubert, *Physical review b* **1999**, 59, 1758.

[56]   M. Gajdoš, K. Hummer, G. Kresse, J. Furthmüller, F. Bechstedt, *Physical Review B—Condensed Matter and Materials Physics* **2006**, 73, 045112.

[57]   H. T. Stokes, S. v. Orden, B. J. Campbell, *Journal of Applied Crystallography* **2016**, 49, 1849.

[58]   B. J. Campbell, H. T. Stokes, D. E. Tanner, D. M. Hatch, *Journal of Applied Crystallography* **2006**, 39, 607.

Table of Contents

A chiral multiferroic phase with simultaneous ferroaxial and ferroelectric order was stabilized under biaxial tensile strains in single-crystals of quasi-1D $P6_3$-BaTiS$_3$. Atomic resolution STEM was used to directly map the two order parameters and reveal the presence of nanoscale chiral domains. Theoretical calculations suggest the presence of a strong coupling between the two order parameters, which can enable electric field control over chirality.

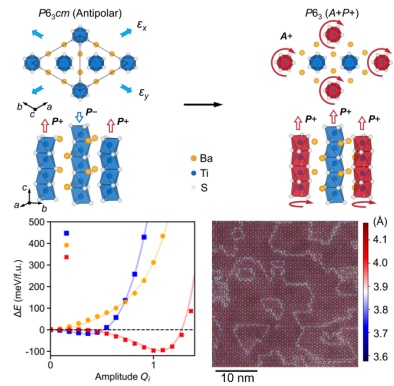